\documentclass[letter]{aa}

  \usepackage{amscd}
  \usepackage{amsmath}
  \usepackage{amssymb}
  \usepackage[T1]{fontenc}
  \usepackage[varg]{txfonts}
  \usepackage{graphicx}
  \usepackage{natbib}
  \usepackage{verbatim}
  \usepackage{array}
  \usepackage{graphicx}

\begin{document}

\title{Off-center explosions of Chandrasekhar-mass white dwarfs: an
  explanation of super-bright type Ia supernovae?}

\author{W. Hillebrandt$^{1}$
        \and 
        S. A. Sim$^{1}$
        \and
        F. K. R\"opke$^{1,2}$}
          
   \offprints{W. Hillebrandt}

   \institute{Max-Planck-Institut f\"ur Astrophysik,
              Karl-Schwarzschild-Str. 1, D-85741 Garching, Germany\\
              \email{[wfh;ssim;fritz]@mpa-garching.mpg.de}
        \and
              Department of Astronomy and Astrophysics, University of
              California, Santa Cruz, 1156 High Street, Santa Cruz CA
              95064, USA
             }
\date{Received / Accepted}

\titlerunning{Off-center explosions and super-bright SNe Ia}
\authorrunning{W.~Hillebrandt, S.~A.~Sim, and F.~K.~R{\"o}pke}

\abstract{The recent discovery of a very bright type Ia supernova,
          SNLS-03D3bb ($\equiv$ SN 2003fg), in the Supernova Legacy
          Survey (SNLS) has raised the question of whether
          super-Chandrasekhar-mass white dwarf stars are needed to
          explain such explosions. In principle, such a progenitor could form
          by the mergers of two rather massive white dwarfs.} 
          {Binary systems of two white dwarfs in close 
          orbit where their total mass significantly exceeds the 
          Chandrasekhar mass, have not yet been found.  Therefore 
          SNLS-03D3bb could
          establish the first clear case of a double-degenerate
          progenitor of a (peculiar) type Ia supernovae. Moreover, if
          this interpretation is correct, it casts some doubt on the
          universality of the calibration relations used to make SNe
          Ia distance indicators for cosmology.}  
          {Here we present a critical discussion of the expected 
          observational finger prints of super-Chandrasekhar-mass 
          explosions; in important respects, these 
          are not consistent with the observations of SNLS-03D3bb.} 
          {We demonstrate that the lop-sided explosion 
          of a Chandrasekhar-mass white dwarf could provide a better 
          explanation.}{}
 
\keywords {Stars: supernovae: general -- Stars: white dwarfs -- Hydrodynamics
          -- Methods: numerical} 

\maketitle


\section{Introduction}

Type Ia supernovae (SNe Ia), i.e. stellar explosions whose spectra show no
hydrogen but lines of elements of intermediate-mass nuclei and of the
iron-group, have received considerable attention recently, mainly because of
their use as distance indicators and as tools to determine cosmological
parameters.

Although the luminosity of nearby, well-observed SNe Ia varies by more
then a factor of ten, empirical correlations between their absolute
magnitude and distance-independent properties, such as the shape of
their light curves, provide means to calibrate their
distances. Several groups have used this approach to estimate the
luminosity of SNe Ia at redshifts up to about 1.5
\citep{riess1998,perlmutter1999,tonry2003,riess2004,astier2006}.
Although they used different calibration procedures and different
distant SN samples, they all agree that SNe Ia at high redshift appear
fainter than they should in a Universe which is matter-dominated
today, suggesting that the Universe started to accelerate its
expansion when it was about half its present age.  This behavior can
be interpreted using cosmologies that include a positive cosmological
constant or a 'dark energy' with negative pressure.

The empirical calibrations can be justified by the fact that they
reduce the scatter in SN Ia Hubble diagrams, but an implicit
assumption in all of them is that all SNe Ia have similar progenitors,
namely almost spherically symmetric white dwarfs at or near the
Chandrasekhar mass, consisting of carbon and oxygen and disrupted by a
thermonuclear combustion wave. In this scenario the (bolometric) peak
luminosity is proportional to the amount of newly synthesized
radioactive $^{56}$Ni (`Arnett's rule', \citet{arnett1982}) and the
observed peak luminosity/ lightcurve shape relation is the result of
secondary effects such as different opacities and/or thermal
structures. It would be difficult to conceive ways to obtain the
observed simple relationships if the progenitors were to span a wide
range of masses, as could be the case for objects formed by the merger
of two white dwarfs. Therefore, the presently favored evolutionary
track to explosion is accretion onto the white dwarf from a
main-sequence or sub-giant companion star rather than a merger.

An obvious prediction of the Chandrasekhar-mass model is a limiting
peak bolometric luminosity of SNe Ia of about $M_{\mbox{\scriptsize
bol}}$ = -20, corresponding to 1.4 M$_\odot$ of $^{56}$Ni, the
Chandrasekhar mass. This is confirmed by a recent parameter
study \citep{woosley2006} which shows that only extreme
Chandrasekhar-mass models can reach B-band magnitudes close to -20,
i.e., models in which it is assumed that almost the entire star is
burned. In reality, of course, the Ni mass has to be lower, allowing
for at least about 0.3 M$_\odot$ in intermediate-mass nuclei, as seen
in the spectra of SNe Ia around maximum light, and taking into
consideration that nuclear burning will produce stable iron-group
nuclei also. In fact, typical deflagration Chandrasekhar-mass
explosion models produce about 0.7 M$_\odot$ or less of radioactive
Ni, and even for a detonation incinerating the entire white dwarf to
iron-group nuclei, some of the ash would be in the form of stable iron
and nickel.

Therefore it came as a great surprise when as part of the SNLS on
April 24, 2003, a SN Ia was discovered at a redshift $z = 0.244$ with
a peak luminosity in the rest frame V band of 20.5 mag, corresponding
to an absolute magnitude of $M_V = -19.94 \pm 0.06$ for a standard
cosmology \citep{howell_et_al2006}.  Using 'Arnett's' rule
\citep{arnett1982}, Howell et al. infer a mass of radioactive Ni of
$(1.29 \pm 0.07)$ M$_\odot$ and conclude that SNLS-03D3bb most likely
was the result of the explosion of a rapidly spinning
super-Chandrasekhar-mass white dwarf. A similar conclusion was reached
by \citet{jeffery2006}. In fact, in their model the mass of the
progenitor has to be about 2 M$_\odot$ in order to account for the 
photospheric velocity of SNLS-03D3bb -- about 8,000 km~s$^{-1}$ on day
+2 after B-maximum (\citealt{howell_et_al2006}; see the discussion in 
the supplementary material to their paper). \citet{howell_et_al2006} 
also mention briefly the possibility of an asymmetric explosion but 
conclude that this effect would be too small to account for the high 
luminosity of SNLS-03D3bb/SN 2003fg. 

In this {\sl Letter} we shall first discuss this scenario in the light
of explosion models of super-Chandrasekhar-mass white dwarfs and will
argue that the photometric and spectroscopic data seem to exclude this
possibility for SNLS-03D3bb. We will then present an alternative
scenario in which nuclear burning in a Chandrasekhar-mass white dwarf
ignites off-center and drives a lop-sided explosions. Here, the high
apparent luminosity of SNLS-03D3bb would result from a nickel blob
rising inside the star towards the observer. Synthetic lightcurves for
a toy model are presented which support our interpretation. In a
forthcoming paper consequences for supernova cosmology will be
discussed.

\section{Super-Chandrasekhar-mass SN Ia models and their predictions}

In principle, a close binary system of two massive white dwarfs
can merge and form a new rapidly spinning degenerate star that
considerably exceeds the Chandrasekhar limit in which the final
configuration is stabilized by rotation. It was shown by
\citet{mueller1985} that such stars can have masses well in excess of
2 M$_\odot$ and angular momenta sufficiently low to avoid the various
rotational instabilities, provided that they rotate
differentially. Additionally, it is not excluded that accretion onto a 
normal white dwarf might have similar results under certain circumstances
\citep{yoon2005}.

If such a star loses angular momentum or the angular momentum is
redistributed closer to rigid rotation, it will become unstable,
contract, and ignite its nuclear fuel. Central ignition and off-center
ignition are conceivable, as well as a spontaneous detonation or a
deflagration.  

\citet{steinmetz1992} have carried out an extended parameter study of
detonations of differentially and rigidly rotating super-Chandrasekhar-mass
C+O white dwarfs in 2D assuming axi-symmetry.  This is a fair approximation
since it was shown by \citet{benz1990} by means of SPH simulations that after
a few revolutions the merging white dwarf pair approaches an almost
axially-symmetric hydrostatic configuration. The hope of \citet{steinmetz1992}
was that due to the lower densities of the rapidly spinning models, a
detonation would find enough white dwarf material at sufficiently low
densities, $\rho \lesssim 10^7$ g~cm$^{-3}$, that carbon and oxygen would
burn to silicon and not to nickel.

To that end, the results of the study were disappointing. The models did
produce amounts of Ni that could easily bring them into agreement with very bright
supernovae such as SN 1991T (or SNLS-03D3bb). However, no matter whether central
or off-center explosions were considered, or rigidly or differentially rotating
progenitors of different mass, the amount of Si produced never exceeded a few 10$^{-2}$
M$_\odot$. This is, by a considerable margin, too little to explain the
strong Si II line seen in SN Ia spectra, including SNLS-03D3bb.  
The reason is simply that the detonation does
not leave enough time for the star to expand and the amount of low-density
progenitor material is small, even in rapidly spinning white dwarfs. Moreover,
for all cases considered the expansion velocities were much higher than those
inferred for SNLS-03D3bb from the spectrum observed at day +2. Consequently, a
prompt detonation of a super-Chandrasekhar-mass white dwarf is clearly ruled
out.

More recently, \citet{pfannes2006} in his thesis has computed models of
both, deflagrations and detonations of super-Chandrasekhar-mass white
dwarfs. The code he used is essentially that of \citet{reinecke2002},
modified and improved by \citet{roepke2005} and \citet{schmidt2006}
for pure deflagrations and by \citet{maier2006} to handle detonation in
C+O fuel. As far as detonations are concerned, Pfannes's work confirms the
earlier results of \citet{steinmetz1992}. Although in a few cases and for
certain rotation laws he finds up to 15\% of the mass is burned to
intermediate-mass nuclei, barely enough to produce the Si II feature
seen in the spectrum of SNLS-03D3bb, the velocity (10,000 to
20,000 km~s$^{-1}$) of this Si would be far too high.

Deflagration models of super-Chandrasekhar-mass white dwarfs, on the
other hand, fall short of producing the amount of Ni required for
bright SNe Ia. Typical differentially rotating
super-Chandrasekhar-mass models predict even less Ni than non-rotating
or rigidly rotating Chandrasekhar-mass models. The reason is easy 
to understand. Differential rotation suppresses the growth of the
Rayleigh-Taylor instability which is the prerequisite for turbulence
and, in turn, determines the rate of fuel consumption. Differential
rotation (needed to increase the mass beyond the Chandrasekhar limit)
thus decreases the burning rate. Finally, the spectra expected
from such models would fit neither those of normal nor of bright SNe Ia
\citep{pfannes2006}.

The only other possible super-Chandrasekhar-mass case, which has not
been investigated until now, is that of a deflagration-to-detonation
transition in a rapidly spinning white dwarf. However, it is not clear
if one would get much more Ni in this case than from a pure deflagration.
The start of such a model would be the deflagration phase discussed
previously. The deflagration-to-detonation transition could occur in
unburned C+O near the center at densities where complete nuclear burning would
take place. But it could also start at low density, thus resulting in
incomplete burning which would produce intermediate-mass nuclei and not the 
iron group. Which of these two possibilities may occur, if either, will
have to be determined from future simulations.

\section{Lop-sided Chandrasekhar-mass explosions}

Here we suggest an alternative model to explain the luminosity
of very bright SNe Ia. The model we propose has two
ingredients: first, it needs about 1 M$_\odot$ of radioactive Ni
and secondly, the radioactive Ni responsible for the optical display 
of the supernova has to be distributed unevenly on large scales. 

\begin{figure}
\includegraphics[width=\linewidth]{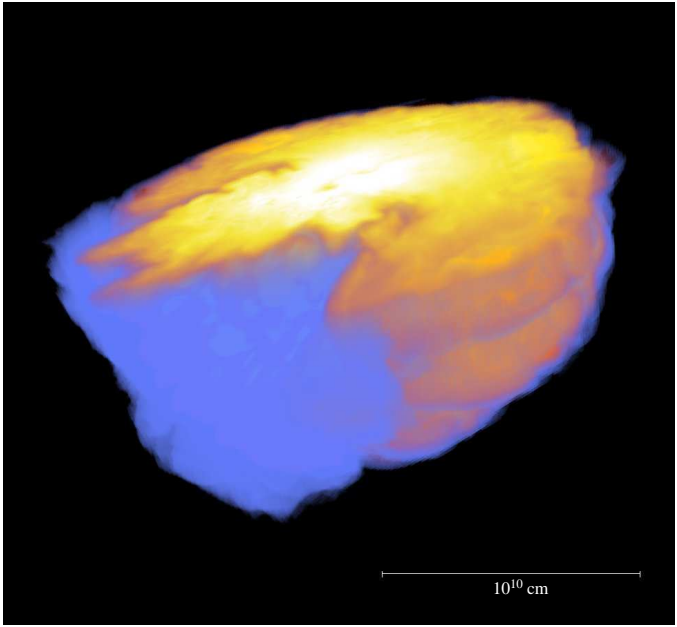}
\caption{Chemical structure of a pure-deflagration off-center 3D
explosion of a Chandrasekhar-mass C+O white dwarf 10 s after
ignition. Half of the star is shown. Volume rendered is the density 
multiplied by the respective mass fractions: iron-group elements
(mostly $^{56}$Ni, shown in yellow/brown); unburned C+O (shown in
blue).
\label{fig1}} 
\end{figure}

Numerical simulations have shown ways in which this could happen in a real
supernova. The rather high mass of Ni could result from a 
deflagration-to-detonation transition (DDT) taking place when 
the star has expanded
to densities around 10$^7$ g cm$^{-3}$ during an initial deflagration
phase \citep{gamezo2005,golombek2005,roepke2007}. The asymmetry of the
Ni distribution could be an intrinsic property of a DDT at one or a
few points \citep{livne1999}, or might be a consequence of
an already lop-sided deflagration \citep{calder2004,roepke2006} (see
Fig.~\ref{fig1}, model 3T2d200 of \citet{roepke2006} as an
example of a significantly aspherical deflagration model; note,
however, that while this deflagration was sufficient to unbind the
star, no DDT was assumed for this particular case). 
In principle,
asymmetric explosions might also result if after an off-center
deflagration phase the white dwarf stays gravitationally bound and a
surface wave triggers a secondary detonation
(\citealp{plewa2004,plewa2006}, but see \citealp{roepke2006}). However, since
`surface-detonation models' like those of
\citet{plewa2006} predict expansion velocities in excess of those
inferred from the observed spectrum of SNLS-03D3bb
\citep{howell_et_al2006}, that particular scenario does not seem
appropriate here.

In order to investigate the observable consequences of a lop-sided
distribution of radioactive nickel in a rather general way, we
have used a class of simply-parameterised models in which a spherical
blob of nickel-rich material is placed in a homologously expanding
spherical supernova. Our models have two free parameters only:
the mass of Ni and its displacement. Guided by the numerical
simulations discussed earlier, we assume that the centre-of-mass of
the nickel blob is displaced from the origin along the $+z$-direction
by an arbitrary velocity shift ($\Delta v$).  Within the nickel-rich
blob, a fixed initial mass fraction of $^{56}$Ni is adopted ($f$); it
is assumed that there is no $^{56}$Ni outside the blob. Although
simplistic, these models provide a convenient and readily
understandable tool to quantify effects.

\begin{figure}
\centerline{\includegraphics[width=\linewidth]{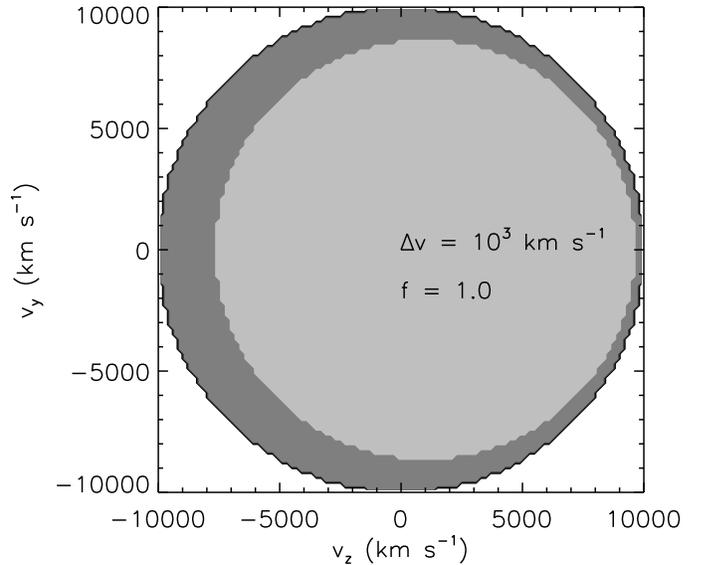}}
\caption{Slice through the $y$-$z$ plane showing the distribution of
$^{56}$Ni adopted in the model. The dark grey area indicates the
regions in which there is no $^{56}$Ni while the light grey area,
the centre of which is displaced by $\Delta v = 10^3$~km~s$^{-1}$ from
the origin, 
contains pure $^{56}$Ni ($f = 1$). 
The models is symmetric
under rotation about the z-axis. \label{fig2}}
\end{figure}

In the context of SNLS-03D3bb, we will discuss one particular
realisation of the model. A slice through this particular geometry 
(in velocity space) is shown in Fig.~\ref{fig2}. The model has a
total mass of 1.4~$M_{\odot}$, an initial $^{56}$Ni-mass of
0.89~$M_{\odot}$, a maximum expansion velocity of $10^4$~km~s$^{-1}$
and uniform mass density. The nickel blob parameters are $f = 1$ and
$\Delta v = 10^3$~km~s$^{-1}$. Wider discussion of these lop-sided
models and their implications will be presented by Sim et al. (in
preparation). 

\begin{figure}
\centerline{\includegraphics[width=\linewidth]{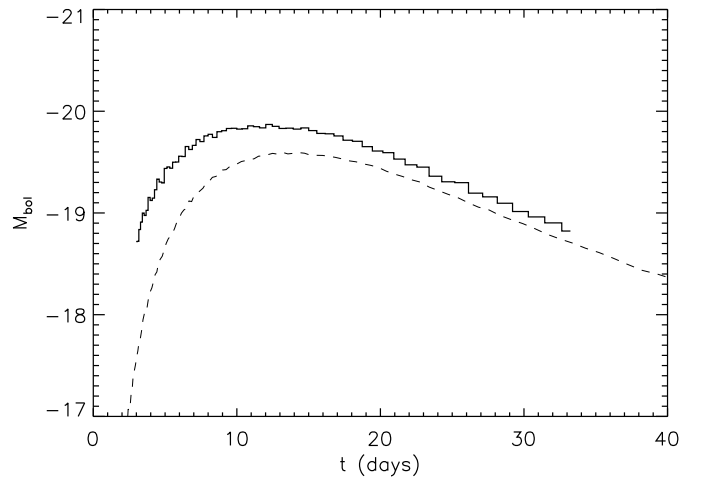}}
\caption{Bolometric light curves computed from the model. The
dashed line is the angle-averaged light curve while the solid histogram 
indicates the light curve as viewed from the $+z$-direction. \label{fig3}}
\end{figure}

Using the Monte Carlo radiative transfer code described by \citet{sim2007},
bolometric lightcurves were computed for various viewing angles to
the model. A uniform, grey-absorption cross-section of
0.1~cm$^{2}$~g$^{-1}$ was adopted; this simple treatment is expected to
maximise the apparent viewing-angle dependence of the light curve and
so allows a limit on the scale of any such effect to be established.
In Fig.~\ref{fig3}, two of the computed light curves are shown: first, the
angle-averaged light curve (dashed line; this is the lightcurve
obtained by averaging over those seen from a large number of randomly
oriented lines-of-sight) and secondly, the light curve recorded by a
distant observer on the $+z$-axis (histogram).

As one would expect, the angle-averaged light curve from the model
cannot account for the observed peak brightness of SNLS-03D3bb; the
peak magnitude is only $\sim -19.58$~mag compared to
$M_{\mbox{\scriptsize bol}} = -19.87 \pm 0.06$~mag as reported by
\citet{howell_et_al2006}. However, the lop-sided nature of the model
causes the light curve to be dependent on viewing angle such that if
viewed along the direction in which the nickel blob is displaced, the
peak is nearly 0.3~mag brighter. This effect brings us to consistency
with the observed peak brightness of SNLS-03D3bb.

A valid question, of course, is whether our model meets the velocity
constraint of SNLS-03D3bb. Of course, we cannot answer
this question for a "real" 3D off-center DDT of the sort we propose
because hydrodynamical simulations for such a model have not yet been 
performed. Most likely the velocity will be similar to
what is obtained in published (less asymmetric) DDT models.
For example, model (a) of \citet{gamezo2005} predicts a maximum
expansion velocity of the outermost layers of about 10,000 km~s$^{-1}$ 
for a total Ni mass of 0.8 $M_{\odot}$. On day +2 (when the spectrum of 
SNLS-03D3bb was taken) and later, most SNe Ia show constant Si velocity,
indicating that the Si-layer is outside the photosphere. For the
Gamezo et el. model, this means that the Si velocity
at that epoch would be $\sim$8,000 km~s$^{-1}$, in fair agreement with 
SNLS-03D3bb.

We can be a little more precise for the toy model we used here.  We
assumed a maximum velocity of 10,000 km~s$^{-1}$ (this is somewhat
arbitrary, however, and in principle we could have adopted a lower
or a higher velocity). As in the DDT model discussed above, such
a peak ejection velocity does not cause a problem because the majority
of the ejecta, even in the narrow region above the Ni-bubble on the
front hemisphere of the model, has lower velocities than the maximum
value. Also, projection effects will always cause the line-of-sight
velocity deduced from the spectrum to be smaller than the true
velocity.  In fact, most of the material in front of the Ni-bubble has
a typical line-of-sight velocity of 6,000 km~s$^{-1}$ only. Thus,
although we cannot make detailed predictions for the spectrum on day
+2 without undertaking considerably more sophisticated radiative
transfer calculations, we conclude that the velocities reported by
\citet{howell_et_al2006} are not inconsistent with our model.

The model considered here is tuned in a number of respects. First, the
combination of Ni mass (0.89~$M_{\odot}$) and geometry
(Fig.~\ref{fig2}) is arranged to produce close to the maximum effect
possible. Further increasing the mass of the Ni blob does not allow a
peak magnitude brighter than $\sim -20$ to be obtained from any
direction; although more Ni makes the angle-averaged lightcurve
brighter, it does this at the expense of suppressing the angular
variation since the mass of the surrounding shell becomes smaller. For
lower Ni masses it is possible to obtain more significant angular
variations but the absolute magnitude rapidly becomes too faint.

Secondly, the treatment of opacity is also extreme. The adopted
cross-section (0.1~cm$^{2}$~g$^{-1}$) may be optimistically high
for material in which significant nuclear burning has not taken place.
Since it is the opacity in the outer regions which leads to the
angular variations in the lightcurve, it is likely that 
even these outer regions much be burned to at least intermediate mass
elements. Conversely, the adopted cross-section may be too low for the
region inside the Ni-rich blob; if the opacity were higher, the
lightcurve would rise more slowly and have a correspondingly dimmer
peak. To investigate the effects of the opacity treatment in detail
would require much more complex non-grey, non-LTE radiation transport
which lies beyond the scope of this letter but which is vital for
fully understanding these effects in the future.

\section{Conclusions} 

In light of these considerations, we draw the following
conclusions. It may be possible to observe lightcurves with peak
brightness close to that of SNLS-03D3bb/SN 2003fg from supernovae 
with a lop-sided distribution of radioactive material, and the total
mass of such material is only $\geq 0.9~M_{\odot}$; thus this model is
a viable alternative to the proposed super-Chandrasekhar theory.

However, since the conditions on a simple off-centre explosion model
which can explain the brightness of SNLS-03D3bb are quite tight, we
can make several predictions from this hypothesis which may be
testable via future observations. First, such objects must be rare
since both a moderately high nickel mass ($\geq 0.9~M_{\odot}$) and
rather special viewing direction are required. Secondly, SNLS-03D3bb
lies very close to the limit of brightness that can be readily
achieved -- thus any supernovae more than one or two tenths of a
magnitude brighter could not be explained in this way.  Thirdly, the
asymmetries postulated in our model should leave markers in the
nebular spectra, such as a net redshift of the lines of
intermediate-mass elements.  And finally, our calculations suggest
that extremely bright SNe~Ia should have shorter-than-average
light-curve rise times.



\begin{thebibliography}{26}
\expandafter\ifx\csname natexlab\endcsname\relax\def\natexlab#1{#1}\fi

\bibitem[{{Astier} {et al.}(2006)}]{astier2006}
{Astier}, P., {Guy}, J., {Regnault}, N., {et al.}, Astron. \ \& Astrophys., 
447, 31

\bibitem[{{Arnett}(1982)}]{arnett1982}
{Arnett}, W.D. 1982, Astrophys.\ J., 253, 785

\bibitem[{{Benz} {et~al.}(1990)}]{benz1990}
{Benz}, W., {Cameron}, A.G.W., {Press} W.H., \& {Bowers}, R.L. 1990,
Astrophys.\ J., 348, 647

\bibitem[{{Calder} {et~al.}(2004)}]{calder2004}
{Calder}, A.~C., {Plewa}, T., {Vladimirova}, N., {Lamb}, D.~Q., \& {Truran},
  J.~W. 2004, astro-ph/0405126

\bibitem[{{Gamezo} {et~al.}(2005)}]{gamezo2005}
{Gamezo}, V. N., {Khokhlov}, A. M., \& {Oran}, E. S. 2005,
Astrophs.\ J., 623, 337


\bibitem[{{Golombek} \& {Niemeyer}(2005)}]{golombek2005}
{Golombek}, I \& Niemeyer, J.~C. 2005, Astron.\ \& Astrophys., 438, 611 

\bibitem[{{Howell} {et~al.}(2006)}]{howell_et_al2006}
{Howell}, D.A., {Sullivan}, M., {Nugent}, P., {et al.} 2006, Nature, 
443, 309  


\bibitem[{{Jeffery} {et al.}(2006)}]{jeffery2006}
{Jeffery}, D.~J., {Branch}, D., \& {Baron}, E. 2006, astro-ph/0609804 


\bibitem[{{Livne}(1999)}]{livne1999}
{Livne}, E. 1999, Astrophs.\ J., 527, L97

\bibitem[{{Maier} \& {Niemeyer}(2006)}]{maier2006}
{Maier}, A. \& {Niemeyer}, J.C. 2006,  Astron.\ \&
  Astrophys., 451, 207

\bibitem[{{M\"uller} \& {Eriguchi}(1985)}]{mueller1985}
 {M\"uller}, E. \& {Eriguchi}, Y. 1985, Astron.\ \&
  Astrophys., 152, 325

\bibitem[{{Perlmutter} {et~al.}(1999)}]{perlmutter1999}
{Perlmutter}, S., {Aldering}, G., {Goldhaber}, G., {et~al.} 1999, 
Astrophys.\ J., 517, 565

\bibitem[{{Pfannes}(2006)}]{pfannes2006}
{Pfannes}, J.M.M. 2006, PhD Thesis, Universit\"at W\"urzburg; 

\bibitem[{{Plewa} {et~al.}(2004)}]{plewa2004}
{Plewa}, T., {Calder}, A.~C., \& {Lamb}, D.~Q. 2004, Astrophys.\ J., 
612, L37

\bibitem[{{Plewa}(2006)}]{plewa2006}
{Plewa}, T., 2006, astro-ph/0611776, Astrophys.\ J., in press

\bibitem[{{Reinecke} {et~al.}(2002)}]{reinecke2002}
{Reinecke}, M., {Hillebrandt}, W., \& {Niemeyer}, J.~C. 2002, Astron.\ \&
  Astrophys., 386, 936

\bibitem[{{Riess} {et~al.}(1998)}]{riess1998}
{Riess}, A.~G., {Filippenko}, A.~V., {Challis}, P., {et~al.} 1998, Astron. J.,
  116, 1009

\bibitem[{{Riess} {et al.}(2004)}]{riess2004}
{Riess}, A.~G., {Strolger}, L.-G., {Tonry}, J., {et al.}, Astrophys.\ J., 
607, 665 

\bibitem[{{R{\"o}pke}(2005)}]{roepke2005}
{R{\" o}pke}, F.~K. 2005, Astron.\ \& Astrophys., 432, 969

\bibitem[{{R{\"o}pke} \& {Niemeyer}(2007)}]{roepke2007}
{R{\"o}pke}, F.~K. \& Niemeyer, J.~C. 2007, Astron.\ \& Astrophys., in press

\bibitem[{{R{\"o}pke} {et~al.}(2006)}]{roepke2006}
{R{\" o}pke}, F.~K.,  {Woosley}, S.~E., \& {Hillebrandt}, W.,
2006, astro-ph/0609088, subm. to  the Astrophys.\ J.

\bibitem[{{Schmidt} {et~al.}(2006)}]{schmidt2006}
{Schmidt}, W., {Niemeyer}, J.C., {Hillebrandt}, W., \& {R{\" o}pke}, F.~K.
2006,  Astron.\ \&  Astrophys., 450, 283

\bibitem[{{Sim} (2007)}]{sim2007}
{Sim}, S. A., 2007, MNRAS, in press

\bibitem[{{Steinmetz} {et~al.}(1992)}]{steinmetz1992}
{Steinmetz}, M., {M\"uller}, E. \& {Hillebrandt}, W. 1992, Astron.\ \&
  Astrophys., 254, 177

\bibitem[{{Tonry} {et al.}(2003)}]{tonry2003}
{Tonry}, J.~L., {Schmidt}, B. P., {Barris} B., {et al.}, Astrophys.\ J., 
594, 1


\bibitem[{{Woosley} {et~al.}(2006)}]{woosley2006}
{Woosley}, S.~E., {Kasen}, D., {Blinnikov}, S., \& {Sorokina}, E. 2006,
astro-ph/0609562, subm. to  the Astrophys.\ J.


\bibitem[{{Yoon} \& {Langer}(2005)}]{yoon2005}
{Yoon}, S.-C., \& {Langer}, N. 2005, Astron.\ \&  Astrophys., 435, 967 

\end{thebibliography}
\end{document}